\documentclass[aps,prl,twocolumn,longbibliography,superscriptaddress]{revtex4-2}
\usepackage{bbm}
\usepackage{graphicx}
\usepackage{dcolumn}
\usepackage{bm}
\usepackage{subfigure}
\usepackage{amsmath}
\usepackage{feynmf}
\usepackage{hyperref}
\usepackage{amssymb}
\usepackage{attachfile}
\usepackage{multirow}
\usepackage{makecell}
\usepackage{pifont}
\usepackage{soul}
\usepackage{cancel}

\newcommand{\bk}{\boldsymbol k}

\newcommand{\bq}{\boldsymbol q}
\newcommand{\bQ}{\boldsymbol Q}

\newcommand{\bd}{\boldsymbol d}
\newcommand{\ba}{\boldsymbol a}

\newcommand{\bA}{\boldsymbol{A}}

\newcommand{\zb}{\color {black}}

\usepackage{braket}
\usepackage{esint}
\usepackage{times}

\begin{document}

\title{Mixed-Parity Altermagnetism in Collinear Spin-Orbital Magnets}

\author{Zheng-Yang Zhuang}
\email{zhuangzhy3@mail2.sysu.edu.cn}
\affiliation{Guangdong Provincial Key Laboratory of Magnetoelectric Physics and Devices,
	State Key Laboratory of Optoelectronic Materials and Technologies,
	School of Physics, Sun Yat-sen University, Guangzhou 510275, China}
	
\author{Jin-Xin Hu}
\affiliation{Department of Physics, The Hong Kong University of Science and Technology, Clear Water Bay, Hong Kong, China}

\author{Song-Bo Zhang}
\email{songbozhang@ustc.edu.cn}
\affiliation{Hefei National Laboratory, Hefei, 230088, China}
\affiliation{School of Emerging Technology, University of Science and Technology of China, Hefei, 230026, China}

\author{Lun-Hui Hu}
\email{lunhui@zju.edu.cn}
\affiliation{Center for Correlated Matter and School of Physics, Zhejiang University, Hangzhou 310058, China}

\author{Zhongbo Yan}
\email{yanzhb5@mail.sysu.edu.cn}
\affiliation{Guangdong Provincial Key Laboratory of Magnetoelectric Physics and Devices,
	State Key Laboratory of Optoelectronic Materials and Technologies,
	School of Physics, Sun Yat-sen University, Guangzhou 510275, China}

\date{\today}

\begin{abstract}
	Altermagnetism has so far mainly been understood in its even- and odd-parity forms. We show that collinear antiferromagnets with zero net magnetization can also host mixed-parity spin splitting, namely neither purely even nor purely odd in momentum. We identify the symmetry conditions for such mixed-parity altermagnetism and show that, in two dimensions, it can arise in spin-orbital magnets when the two antiparallel spin sectors are related by a single mirror symmetry. Using a two-sublattice two-orbital model, we demonstrate that circularly polarized light induces mixed-parity altermagnetism at finite staggered potential and odd-parity spin-orbital {\zb altermagnetsim} at zero staggered potential. Mixed-parity altermagnetism thereby emerges as the intermediate spin-split regime between even- and odd-parity {\zb altermagnetism} when spin splitting and zero net magnetization are maintained. Spin-resolved orbital Edelstein effects provide a complementary electrical probe of the underlying spin-orbital order.
\end{abstract}

\maketitle

Nonrelativistic momentum-dependent spin splitting~\cite{Hirsch1990,Ikeda1998,wu2004,wu2007,Hayami2019NRSS,Hayami2020NRSS,Liu2025} has emerged as a central concept in unconventional magnetism, reshaping our understanding of spin-split electronic structures in antiferromagnets~\cite{Hayami2019NRSS,Hayami2020NRSS,Liu2025}. Among various magnetic structures~\cite{Chen2014PRL,nakatsuji2015large,Hayami2020NRSS,Libor2022review,Hellenes2023pwave,Brekke2024pwave,Zhu2024,Yu2025Odd,Song2025Odd,yamada2025pwave,Liu2025,zhuang2026newton}, altermagnetism (AM) provides a particularly intriguing realization~\cite{Ma2021AM,Libor2022AMb,Libor2022AMa}, as it supports spin-split bands despite vanishing net spin magnetization in a collinear setting~\cite{Yuan2020AM,Libor2022AMa,Yuan2021AM}. Within spin-space group theory~\cite{Liu2022AM,Libor2022AMa,Liu2024AMPRX,Jiang2024SSG,Xiao2024SSG,liu2026OSSG}, altermagnetism was first recognized in its even-parity form~\cite{Ma2021AM,Libor2022AMa} (e.g., $d$-, $g$-, and $i$-wave), where $E_{\bk,s}=E_{-\bk,s}$ with momentum $\bk$ and spin index $s$. Such even-parity altermagnetism has been experimentally identified in diverse materials~\cite{Osumi2024MnTe,Lee2024MnTe,Krempasky2024,Hajlaoui2024AM,Reimers2024,Ding2024CrSb,Yang2024CrSb,Zeng2024CrSb,Li2024CrSb,Lu2024AM7,Jiang2024KV2Se2O,Zhang2025Cpair} and predicted to underlie a wide range of phenomena~\cite{Libor2022AMb,Libor2022AMc,Rafael2021AM,Ouassou2023AM,Bai2023AM,Fang2023NHE,Zhu2023TSC,Lu2024AM11,Zhang2024AM,Zhu2024dislocation,Ghorashi2024AM,Jin2024AM11,Han2024AM,Antonenko2024AM,Hu2025NLME,Hu2025Cpair,Duan2025AFMAM,Lin2025AM4,Chen2025AM7}. More recently, this concept has been extended to odd-parity settings~\cite{lin2025,zeng2025,zhuang2025,zeng2025oddb,li2025floquet,zhu2025floquet,huang2025oddparityAM,liu2025floquet,Pan2025odd,li2025odd}, where $E_{\bk,s}=E_{-\bk,-s}$, by breaking the spinless time-reversal symmetry $[\bar{C}_{2}\Vert\mathcal{T}]$ in conventional $\mathcal{PT}$-symmetric antiferromagnets with spin-degenerate bands. However, this picture remains incomplete from the viewpoint of parity classification. In superconductors, mixed-parity pairing~\cite{Frigeri2004mixed,Bauer2004mixed} constitutes an important class of order parameters beyond purely even- and odd-parity pairing. By analogy, spin splitting in altermagnets can, in principle, also be mixed parity, being neither purely even nor purely odd in momentum. Realizing such a mixed-parity case would extend the known even- and odd-parity forms while preserving the defining ingredients of altermagnetism.

\begin{figure}[t]
	\includegraphics[width=\columnwidth]{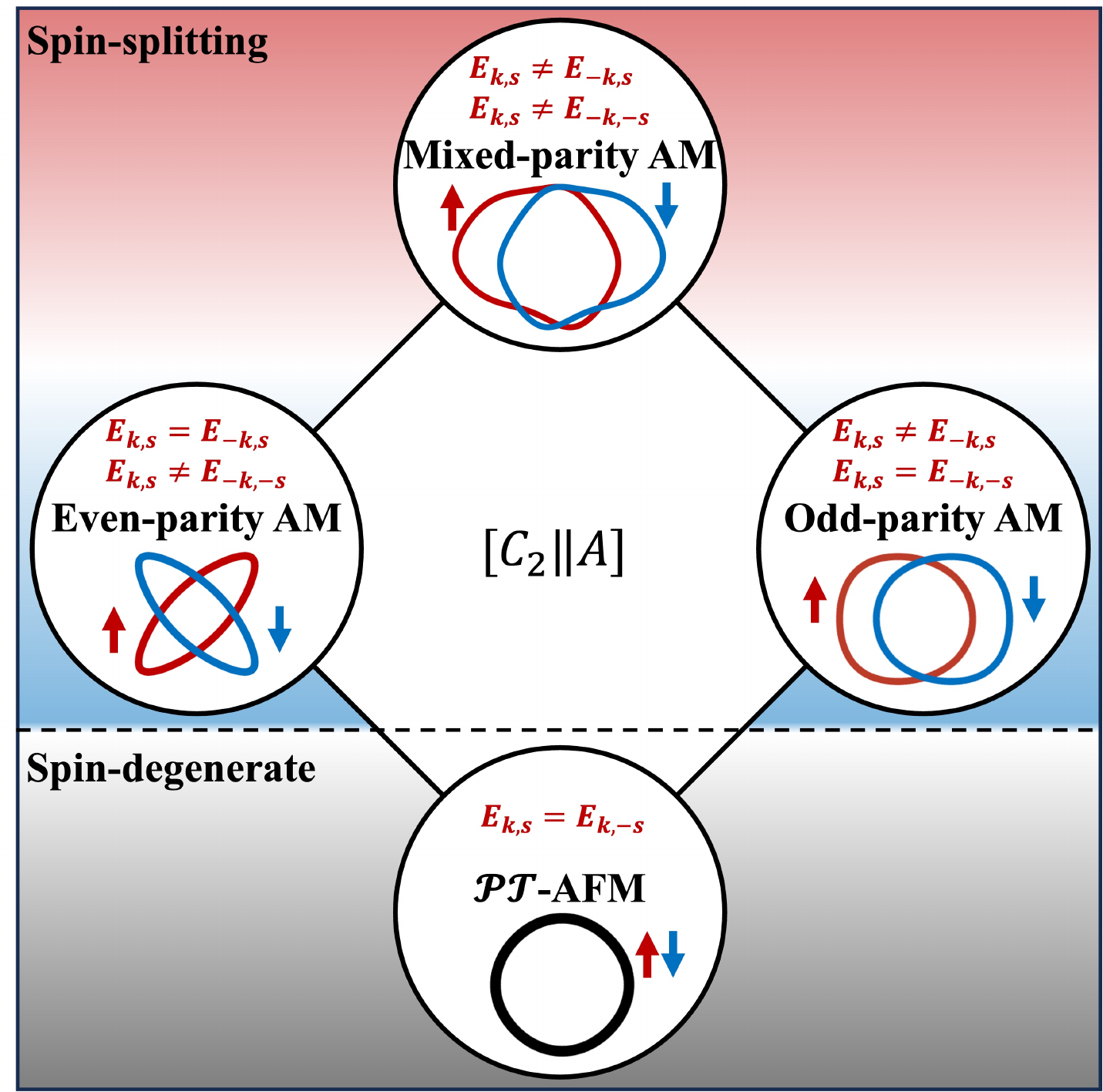}
	\caption{Four classes of collinear antiferromagnets (AFMs) with zero net spin magnetizations. $[C_{2}\Vert A]$ represents the symmetry connecting the two spin sectors.}
	\label{fig:schematic}
\end{figure}

Orbitals, as another fundamental degree of freedom of electrons~\cite{tokura2000orbital}, have been shown to 
{\zb provide a microscopic mechanism for the emergence of} altermagnetism. In particular, when coupled to different sublattices, orbital degrees of freedom can form real-space orbital orders and generate either even-~\cite{Leeb2024AM,Vila2025orbitalorder,Meier2026} or odd-parity~\cite{zhuang2025} altermagnetism. It has also been shown that different orbitals on the same site can couple to opposite spins, giving rise to spin-orbital magnetism~\cite{Giuli2025,Wang2025soAM,Rodrigo2025,Ornellas2026soam}. Unlike relativistic spin-orbit coupling~\cite{Kane2005,Bernevig2006,Galitski2013}, this type of spin-orbital locking breaks time-reversal symmetry and can arise from the interplay between on-site Hubbard and Hund interactions~\cite{lu2025soam}. Importantly, because the spin magnetization is compensated locally on each site, spin-orbital magnetism can retain zero net magnetization even in a two-sublattice system with inequivalent onsite chemical potentials. This is in sharp contrast to Néel-type antiferromagnets with sublattice-staggered local spin moments, where a staggered chemical potential generally induces a finite net spin magnetization and thus drives the system into a ferrimagnetic state~\cite{wang2026cfim}.

In this Letter, we show that collinear spin-orbital magnets provide a natural platform for realizing mixed-parity altermagnetism. 
{\zb We first establish the symmetry conditions required for mixed-parity altermagnetism, then focus on a specific model with spin-orbital orders that preserve lattice translation symmetry. We identify the spin-orbital order capable of realizing mixed-parity altermagnetism and further demonstrate that this state intermediates between even- and odd-parity altermagnetism while preserving spin splitting and zero net magnetization, as illustrated in Fig.~\ref{fig:schematic}}. {\zb Through microscopic calculations, we show that the required spin-orbital order can be interaction-driven. Finally, we suggest that spin-resolved orbital Edelstein effects could provide a probe of this order, complementing spin-resolved ARPES, which directly measures the spin-splitting symmetry}.

{\it \color{blue}Symmetry analysis.---}Mixed-parity altermagnetism is defined by three symmetry conditions. First, as an altermagnetic state, it must host a collinear spin configuration with zero net magnetization~\cite{Libor2022AMa}. Thus, the two antiparallel spin sectors are related by a symmetry of the ordered state. Second, the spin splitting must be non-even-parity, namely $E_{\bk,s}\neq E_{-\bk,s}$ for generic $\bk$. This excludes $[\bar{C}_{2}\Vert\mathcal{T}]$ and, more generally, any symmetry that acts as an inversion operation within a given spin sector~\cite{zhuang2025}. 
For {\zb two-dimensional} systems, out-of-plane rotational symmetries of the form $[C_{2}\Vert C_{nz}]$ with $n=3,4$ are likewise excluded, as they enforce spin-degenerate bands and even-parity spin splitting, respectively. Third, the spin splitting must also be non-odd-parity, namely $E_{\bk,s}\neq E_{-\bk,-s}$ for generic momentum $\bk$. This excludes symmetries that relate opposite spins while reversing momentum, including $[C_{2}\Vert\mathcal{P}]$, effective time-reversal symmetry, and, in {\zb two dimensions (2D)}, rotational symmetries $[C_{2}\Vert C_{nz}]$ with $n=2,6$. More generally, all such forbidden symmetries, whether enforcing even- or odd-parity spin splitting, must remain absent even when combined with operations that leave $\bk$ invariant, such as translation $\tau$ and, in 2D, mirror symmetry $\mathcal{M}_{z}$. Satisfying only the first and second (third) conditions gives odd-parity (even-parity) altermagnetism, while mixed-parity altermagnetism requires all three.

This analysis further shows that, in 2D, a vertical mirror symmetry in real space is singled out among the crystalline symmetries relating the two antiparallel spin sectors. Unlike the rotational, inversion, or effective time-reversal symmetries discussed above, a vertical mirror alone does not enforce purely even- or odd-parity spin splitting. In previously studied even-parity cases, the mirror symmetry ensures zero net spin magnetization, whereas $[\bar{C}_{2}\Vert\mathcal{T}]$ fixes the parity of the spin splitting~\cite{Ma2021AM,Libor2022AMa}. Breaking $[\bar{C}_{2}\Vert\mathcal{T}]$ therefore releases this constraint. This can be achieved by mechanisms that generate orbital magnetization, such as sublattice currents~\cite{lin2025,zeng2025}, chiral orbital order~\cite{zhuang2025}, and dynamical CPL~\cite{huang2025oddparityAM,zhu2025floquet,li2025floquet,liu2025floquet,zhu2026light}. These mechanisms break both $[C_{2}\Vert\mathcal{M}_{i}]$ and $[\bar{C}_{2}\Vert\mathcal{T}]$ while preserving their combination $[C_{2}\Vert\mathcal{M}_{i}\mathcal{T}]$, thereby stabilizing mixed-parity altermagnetism. In this work, we focus on CPL as a highly tunable route to the mixed-parity state.

\begin{figure}[t]
	\includegraphics[width=\columnwidth]{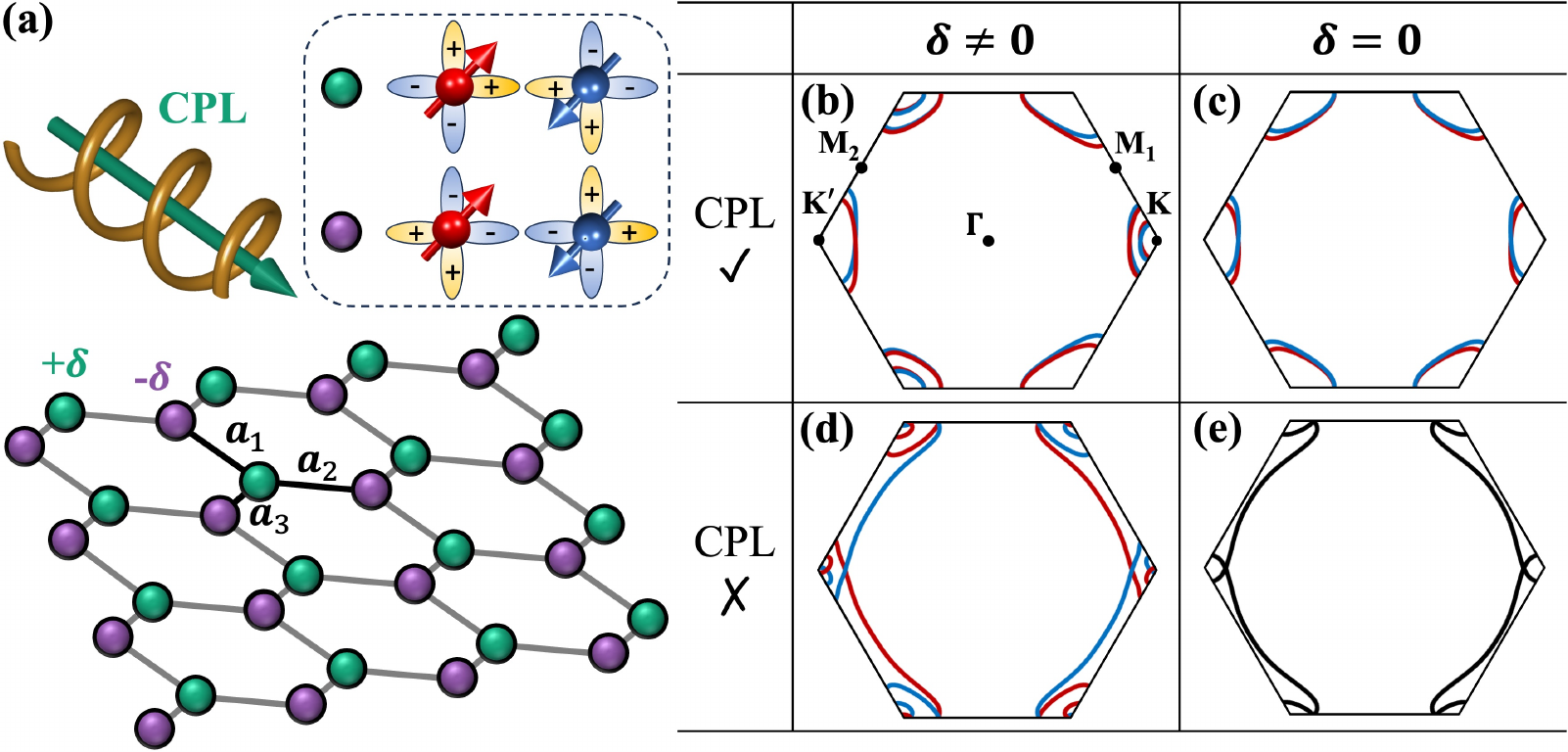}
	\caption{(a) Schematic of the lattice model and the ingredients leading to mixed-parity altermagnetism. The green and purple spheres denote the two sublattices. The dashed box illustrates the $\mathcal{O}_{3}$ spin-orbital order, in which opposite spin sectors on a given site are coupled to different local orbital configurations, represented by the petal-like patterns and spin arrows. The light-yellow and light-blue lobes indicate the sign of the order parameter on the $d_{xz}$ ($x$-oriented lobes) and $d_{yz}$ ($y$-oriented lobes) orbitals. (b-e) Spin-resolved Fermi surfaces at $E=-0.88t_{0}$ [(b-c)] and $E=-1.25t_{0}$ [(d-e)] for the $\mathcal{O}_{3}$ phase. Depending on the presence of $\delta$ and CPL, the spin-resolved Fermi surfaces exhibit (b) mixed-parity, (c) $p_{y}$-wave, (d) $d$-wave, and (e) spin-degenerate patterns. The parameters are $\{t_{0},t_{1},t_{2},t_{3},J_{3},J_{i\neq3}\}=\{1,0.06,0.2,0.1,0.5,0\}$. In panels (b-c), $A_{0}=1$, $\omega=10$. In panels (b,d), $\delta=0.1$.}
	\label{fig:spectrum}
\end{figure}

{\it \color{blue}Spin-orbital magnetism.---}We now implement the above symmetry analysis in a concrete model. To permit non-even-parity spin splitting under driving, we consider a noncentrosymmetric monolayer, thereby excluding inversion symmetry, which would otherwise enforce $E_{\bk,s}=E_{-\bk,s}$.
Within this setting, collinear spin-orbital magnets provide appropriate platforms for realizing mixed-parity altermagnetism. Because opposite spins couple to different real orbitals on the same site, a mirror symmetry that leaves each sublattice invariant while exchanging the two orbitals naturally gives rise to $[C_{2}\Vert\mathcal{M}_{i}]$. Under CPL, its combined antiunitary counterpart $[C_{2}\Vert\mathcal{M}_{i}\mathcal{T}]$ can remain intact and protect mixed-parity altermagnetism. A minimal model therefore consists of two sublattices, two orbitals, and two spins, giving rise to an eight-band Hamiltonian. Without loss of generality, we consider a two-dimensional hexagonal lattice with $d_{xz}$ and $d_{yz}$ orbitals {\zb [see Fig.~\ref{fig:spectrum}(a)]},
\begin{eqnarray}
	\mathcal{H}(\bk)&=&\mathcal{H}_{\rm ele}(\bk)s_{0}+J_{1}\tau_{0}\sigma_{z}s_{z}+J_{2}\tau_{z}\sigma_{z}s_{z}+J_{3}\tau_{x}\sigma_{z}s_{z}\nonumber\\
	&&+J_{4}\tau_{x}\sigma_{0}s_{z}+J_{5}\tau_{z}\sigma_{0}s_{z}+J_{6}\tau_{0}\sigma_{0}s_{z},
	\label{eq: H}
\end{eqnarray}
where $J_{i=1,2,3,4,5,6}$ denotes the order parameters of the different spin-orbital magnetic phases $\mathcal{O}_{i}$, and $\boldsymbol{\tau}$, $\boldsymbol{\sigma}$, and $\boldsymbol{s}$ are Pauli matrices acting on orbital, sublattice, and spin space, respectively. These six channels exhaust the on-site spin-orbital orders compatible with spin conservation {\zb and lattice translation symmetry}. The spin-independent Hamiltonian is
\begin{eqnarray}
	\mathcal{H}_{\rm ele}(\bk)=\left[
	\begin{array}{cc}
		\Delta_{tra}(\bk)+\delta & \Delta_{ter}(\bk)\\
		\Delta^{\dagger}_{ter}(\bk) & \Delta_{tra}(\bk)-\delta
	\end{array}
	\right],
\end{eqnarray}
where $\Delta_{ter(tra)}(\bk)=\sum_{j}e^{-i\bk\cdot\bd_{j}}T_{ter(tra)}(\theta_{j})$ describe the (next-)nearest-neighbor hopping along bond $\bd_{j}$ with polar angle $\theta_{j}$. The corresponding hopping matrices are
$
T_{ter(tra)}(\theta_{j})=t_{0(2)}\tau_{0}+t_{1(3)}\bigl(\cos2\theta_{j}\tau_{z}+\sin2\theta_{j}\tau_{x}\bigr).
$
Here $t_{0}$ ($t_{2}$) and $t_{1}$ ($t_{3}$) denote the intra-orbital and inter-orbital hopping amplitudes for (next-)nearest-neighbor bonds, respectively, while $\delta$ is a staggered sublattice potential. 
The latter can arise from ferroelectric polarization and be controlled by a vertical electric field $E_{z}$~\cite{wang2026cfim}.
For simplicity, we set $|\ba_{i}|=1$ throughout.

Before the drive is applied, the Hamiltonian supports either $d$-wave spin splitting or spin-degenerate bands, depending on the order parameter $J_i$ and the staggered potential $\delta$; the corresponding symmetries and phases are summarized in Table~\ref{tab:sym}. We first consider $\delta=0$. In this limit, the $\mathcal{O}_{4}$ phase realizes the recently discussed intrinsic spin-orbital $d$-wave altermagnetism~\cite{Wang2025soAM}, also referred to as atomic altermagnetism~\cite{Rodrigo2025}, protected by $[C_{2}\Vert\mathcal{M}_{x}]$, which relates opposite spin sectors through an on-site interchange of the orbital combinations $d_{xz}\pm d_{yz}$. The $\mathcal{O}_{1,2,3}$ phases remain spin degenerate because the two spin sectors are related by inversion. By contrast, in the $\mathcal{O}_{5,6}$ phases no symmetry relates the {\zb two} antiparallel spin sectors, and the system enters either a fully compensated ferrimagnetic {\zb state}~\cite{Liu2025fFIM,Feng2025FIM} or a ferromagnetic state.

When $\delta\neq0$, the two sublattices become inequivalent, and phases whose zero magnetization relies on a sublattice-exchanging symmetry generally acquire a finite magnetization. Accordingly, the $\mathcal{O}_{1}$ and $\mathcal{O}_{2}$ phases become ferrimagnets. By contrast, zero net spin magnetization is preserved in the $\mathcal{O}_{3}$ and $\mathcal{O}_{4}$ phases by $[C_{2}\Vert\mathcal{M}_{x}]$, which remains valid at finite $\delta$. Since finite $\delta$ further breaks $[C_{2}\Vert\mathcal{P}]$, the $\mathcal{O}_{3}$ phase becomes a $d$-wave altermagnet. Its $d$-wave form reverses with the sign of $\delta$, enabling electrical control of the spin-splitting pattern, for example through gating. This behavior is qualitatively different from altermagnets with Néel-type order, where a staggered potential typically induces ferrimagnetism.

\begin{table}[t]
	\centering
	\resizebox{\columnwidth}{!}{
		\begin{tabular}{c|cccc|cccc}
			\hline\hline
			&\multicolumn{4}{c|}{Symmetry}&\multicolumn{4}{c}{Spin-splitting}\\\hline
			\multirow{2}{*}{\makecell{AFM\\order}} & \multirow{2}{*}{$\mathcal{P}$} & \multirow{2}{*}{$[C_{2}\Vert\mathcal{P}]$} & \multirow{2}{*}{$[C_{2}\Vert\mathcal{M}_{x}]$} & \multirow{2}{*}{$[C_{2}\Vert\mathcal{M}_{y}]$}&\multirow{2}{*}{\makecell{$\delta$ \checkmark\\CPL \ding{55}}}&\multirow{2}{*}{\makecell{$\delta$ \checkmark\\CPL \checkmark}}&\multirow{2}{*}{\makecell{$\delta$ \ding{55}\\CPL \checkmark}}&\multirow{2}{*}{\makecell{$\delta$ \ding{55}\\CPL \ding{55}}}\\&&&&&&&&\\\hline
			$\mathcal{O}_{1}$& \ding{55} & \checkmark & \ding{55}&\checkmark&ferri.&ferri.&$f$&$\mathcal{PT}$\\\hline
			$\mathcal{O}_{2}$& \ding{55} & \checkmark & \ding{55}&\checkmark&ferri.&ferri.&$p_{x}$& $\mathcal{PT}$\\\hline
			$\mathcal{O}_{3}$& \ding{55} & \checkmark & \checkmark&\ding{55}&$d$& mixed & $p_{y}$& $\mathcal{PT}$\\\hline
			$\mathcal{O}_{4}$& \checkmark & \ding{55}  & \checkmark&\checkmark&$d$&mixed&\multicolumn{2}{c}{$d$}\\\hline
			$\mathcal{O}_{5}$& \checkmark & \ding{55} & \ding{55}&\ding{55}&\multicolumn{4}{c}{ferri.}\\\hline
			$\mathcal{O}_{6}$ & \checkmark & \ding{55} & \ding{55}&\ding{55}&\multicolumn{4}{c}{ferro.}\\\hline
		\end{tabular}
	}
	\caption{Symmetry classification and properties of six spin-orbital orders, with or without sublattice-staggered potential $\delta$ and the presence of CPL. The spin-space group operator $g$ acts on ${\mathcal O}_i$ as follows: $g[{\mathcal O}_i]g^\dagger = \chi_i {\cal O}_i$, where the character $\chi_i=\pm 1$ corresponds to the symbols \checkmark and \ding{55}, respectively. Here, mixed, ferri., ferro., $d$, $f$, $p_{x/y}$, and $\mathcal{PT}$ denote the mixed-parity altermagnetism, ferrimagnetism, ferromagnetism, $d$-, $f$-, $p_{x/y}$-wave altermagnetism, and $\mathcal{PT}$-symmetric antiferromagnetism, respectively.
	}
	\label{tab:sym}
\end{table}

{\it \color{blue}Light-driven mixed-parity altermagnetism.---}Mixed-parity altermagnetism arises naturally in the $\mathcal{O}_{3}$ and $\mathcal{O}_{4}$ phases when a finite sublattice-staggered potential $\delta$ is combined with CPL. We describe the drive by a time-periodic vector potential $\bA(t)=\bA(t+T)$ with period $T=2\pi/\omega$, which enters the Hamiltonian through the Peierls phase $e^{i\bA(t)\cdot\bd_{j}}$ acquired by hopping along bond $\bd_{j}$. In the high-frequency off-resonant regime, where $\hbar\omega$ is the largest energy scale, the periodically driven system is described by the effective Floquet Hamiltonian~\cite{Kitagawa2011,Goldman2014}
\begin{eqnarray}
	\mathcal{H}_{\rm eff}(\bk)=\mathcal{H}_{0}(\bk)+\sum_{n\geq1}\frac{[\mathcal{H}_{n}(\bk),\mathcal{H}_{-n}(\bk)]}{n\omega}+O(\frac{1}{\omega^2}),
\end{eqnarray}
where $\mathcal{H}_{n}(\bk)=T^{-1}\int_{0}^{T}dt\,e^{in\omega t}\mathcal{H}(\bk,t)$ is the $n$th Floquet component. For CPL propagating along the $z$ direction, we take $\bA(t)=A_{0}(\eta\sin\omega t,\cos\omega t,0)$, where $\eta=\pm1$ labels the light helicity. The zeroth-order Floquet Hamiltonian simply renormalizes the hopping amplitudes, namely $\mathcal{H}_{0}=\mathcal{H}[\bk;T_{ter/tra}(\theta_j)\to T_{ter/tra}(\theta_j)J_0(A_{0}|\bd_{j}|)]$, and therefore preserves the symmetries of the undriven Hamiltonian. Here $J_n(x)$ is the $n$th-order Bessel function of the first kind. 
The conversion {\zb of spin-splitting parity} is thus controlled by the leading Floquet correction $\mathcal{H}'(\bk)\equiv[\mathcal{H}_{1}(\bk),\mathcal{H}_{-1}(\bk)]/\omega$.

As shown in Fig.~\ref{fig:spectrum}, the Fermi surfaces in the $\mathcal{O}_{3}$ phase exhibit mixed-parity spin splitting when both $\delta\neq0$ and CPL are present. Since the spin splitting is symmetry-dictated and the intra-sublattice hoppings $\{t_{2},t_{3}\}$ do not alter the essential symmetries, we neglect them for analytical transparency. The leading Floquet correction then takes the form
\begin{eqnarray}
	\omega\mathcal{H}'(\bk)&=&-\sqrt{3}\eta J_{1}^{2}\!\left(A_{0}\right)
	[f_{1}(\bk)\sigma_{z}+f_{2}(\bk)\tau_{z}\sigma_{z}\nonumber\\
	&&+f_{3}(\bk)\tau_{y}+f_{4}(\bk)\tau_{x}\sigma_{z}],
\end{eqnarray}
where the explicit functions $f_i(\bk)$ are given in Supplemental Material (SM) Sec.~I~\cite{supplemental}. 
This correction breaks both $[\bar{C}_{2}\Vert\mathcal{T}]$ and the mirror symmetries $[C_{2}\Vert\mathcal{M}_{i}]$ ($i=x,y$), while preserving $[C_{2}\Vert\mathcal{M}_{x}\mathcal{T}]$. Consequently, the $\mathcal{O}_{3}$ and $\mathcal{O}_{4}$ phases at finite $\delta$ realize mixed-parity altermagnetism. 
Moreover, because $\mathcal{H}'(\bk)\propto\eta$, the Floquet correction is helicity controlled and can shift the Fermi surfaces between valleys, suggesting a route toward valley-polarized spin transport~\cite{Xiao2010review,Schaibley2016Valley,Vitale2018Valley}.

When $\delta=0$, CPL removes $[\bar{C}_{2}\Vert\mathcal{T}]$, while $[C_{2}\Vert\mathcal{P}]$ continues to relate the antiparallel spin sectors. The driven $\mathcal{O}_{1}$, $\mathcal{O}_{2}$, and $\mathcal{O}_{3}$ phases therefore become $f$-, $p_{x}$-, and $p_{y}$-wave altermagnets, respectively. In particular, the $p_{x}$-wave ($p_{y}$-wave) pattern in the $\mathcal{O}_{2}$ ($\mathcal{O}_{3}$) phase is selected by the remaining $[C_{2}\Vert\mathcal{M}_{y}\mathcal{T}]$ ($[C_{2}\Vert\mathcal{M}_{x}\mathcal{T}]$) symmetry. By contrast, the $\mathcal{O}_{4}$ phase remains a $d$-wave altermagnet, protected by the two intact orthogonal effective mirrors $[C_{2}\Vert\mathcal{M}_{i}\mathcal{T}]$ ($i=x,y$). Taken together with the structure of the order parameters $J_i$, these results indicate that spin-orbital altermagnets generally cannot realize 
{\zb an odd-parity phase if the spin-orbit order is uniform across the two sublattices (referring to these $\sigma_{0}$-orders, including
$\mathcal{O}_{4,5,6}$).
The reason is that these orders preserve the inversion symmetry (see Table~\ref{tab:sym}), which enforces even-parity spin splitting. Sublattice-staggered
spin-orbit order} is therefore essential for realizing odd-parity spin-orbital altermagnetism.

The above analysis further shows that mixed-parity altermagnetism is not a fine-tuned exception, but the generic intermediate spin-split phase between even- and odd-parity altermagnets. {\zb  In Fig.~\ref{fig:spectrum}, we illustrate this using the 
$\mathcal{O}_{3}$ phase as an example.} When $\delta=0$ and CPL is absent, it is an antiferromagnet with spin-degenerate bands {\zb [Fig.~\ref{fig:spectrum}(e)]}. A finite $\delta$ generates $d$-wave spin splitting {\zb [Fig.~\ref{fig:spectrum}(d)]}, whereas CPL at $\delta=0$ generates $p$-wave spin splitting {\zb [Fig.~\ref{fig:spectrum}(c)]}. When both are present, the band structure necessarily becomes mixed-parity {\zb [Fig.~\ref{fig:spectrum}(b)]}. In this sense, mixed-parity altermagnetism plays a role parallel to that of $\mathcal{PT}$-symmetric antiferromagnets, which serve as the intermediate phase when spin degeneracy is allowed. This follows from symmetry: the symmetry groups enforcing even- and odd-parity spin splitting are mutually incompatible, and if both were present simultaneously, the bands would become spin degenerate. Therefore, when tuning between even- and odd-parity altermagnets while maintaining both spin splitting and zero net magnetization, mixed-parity altermagnetism naturally emerges as the intermediate regime.

\begin{figure}[t]
	\includegraphics[width=\columnwidth]{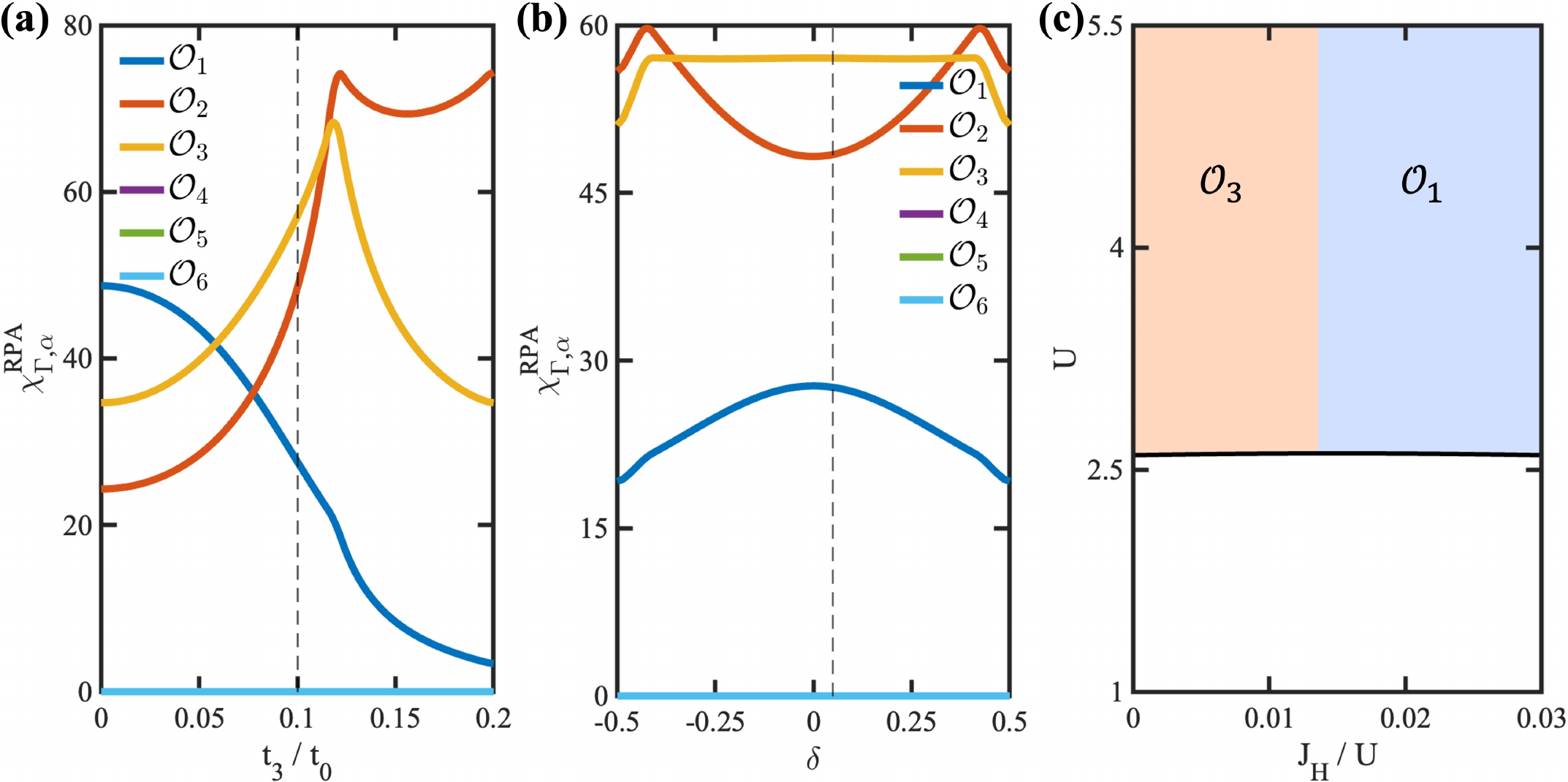}
	\caption{$\chi_{\rm spin}^{\rm RPA}(0,0)$ at $\{U,J_{H}/U\}=\{0.99U_{c},0.001\}$ as a function of (a) the parameter $t_{3}/t_{0}$ and (b) $\delta$ for the six channels, with the dashed line marking the parameters $t_{3}/t_{0}$ and $\delta$ adopted in (c), which represents the $U$-$J_{H}/U$ phase diagram. $U$ and $J_{H}$ denote the intra-orbital Hubbard interaction and Hund coupling, respectively. Parameters are $\{t_{0},t_{1},t_{2},\mu\}=\{1,0.06,0.2,-0.5\}$. In (a), we set $\delta=0.05$, in (b), we set $t_{3}/t_{0}=0.1$.}
	\label{fig:rpa_phase_diagram}
\end{figure}

{\it \color{blue}Magnetic phase diagram.---}We next establish the microscopic accessibility of the spin-orbital orders introduced above. Starting from a repulsive two-orbital Hubbard-Hund model, we analyze magnetic instabilities using the multi-orbital random-phase approximation (RPA)~\cite{lu2025soam,supplemental},
$
	\chi_{\rm spin}^{\rm RPA}(\bq)=\bigl[I-\chi^{(0)}(\bq)\mathcal{U}_{s}\bigr]^{-1}\chi^{(0)}(\bq),
$
where $\chi^{(0)}$ is the bare susceptibility and $\mathcal{U}_{s}$ the spin-channel interaction vertex~\cite{supplemental}. A divergence at the critical interaction $U_{c}$ signals a magnetic instability with ordering wave vector $\bQ$. We focus on $\bQ=0$, where the ordered state preserves the lattice periodicity and can be classified into the six spin-orbital channels $\mathcal{O}_{1,2,3,4,5,6}$. The leading instability is then identified by projecting $\chi_{\rm spin}^{\rm RPA}$ onto these channels and comparing the projected susceptibilities $\chi_{\Gamma,\alpha}^{\rm RPA}$~\cite{supplemental}.

{\zb Figure}~\ref{fig:rpa_phase_diagram} presents the projected RPA susceptibilities versus $t_{3}/t_{0}$ and $\delta$, together with the $U$-$J_{H}/U$ phase diagram at the representative parameters marked by dashed lines. The leading instability is governed mainly by the competing $\mathcal{O}_{1}$, $\mathcal{O}_{2}$, and $\mathcal{O}_{3}$ channels. Increasing $t_{3}/t_{0}$ drives the dominant channel from the Néel-like $\mathcal{O}_{1}$ order to $\mathcal{O}_{3}$, and then to $\mathcal{O}_{2}$ at larger $t_{3}/t_{0}$. The $\mathcal{O}_{3}$ instability remains stable over a broad range of $\delta$, while large $|\delta|$ favors $\mathcal{O}_{2}$. The resulting phase diagram contains a sizable $\mathcal{O}_{3}$ region at small $J_{H}/U$, whereas $\mathcal{O}_{1}$ dominates at larger $J_{H}/U$. Its robustness against chemical potential is shown in the SM Sec.II~\cite{supplemental}. Thus, provided off-resonant light does not substantially reconstruct the spin-orbital order, the $\mathcal{O}_{3}$ phase serves as a realistic parent state for the light-driven odd-parity phase at $\delta=0$ and the mixed-parity phase at $\delta\neq0$.

\begin{figure}[t]
	\includegraphics[width=\columnwidth]{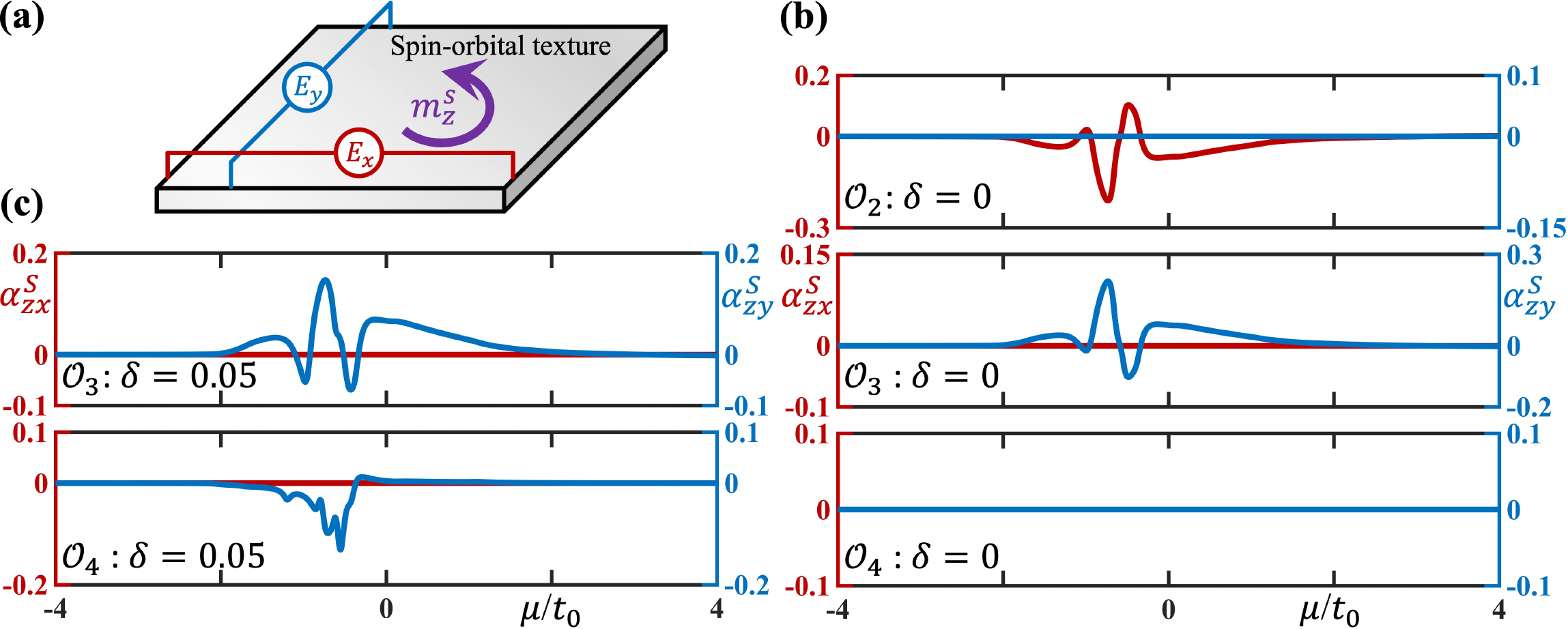}
	\caption{(a) Schematic of the SOEE. In spin-orbital magnets with spin-orbital texture, an electric field $E_{i=x,y}$can induce an orbital polarization $m_{i}^{s}$ in each spin sector. (b)-(c) SOEE (in units of $e^2\tau/(2\hbar^2)$) as a function of $\mu/t_{0}$ before CPL is applied for three representative channels, $\mathcal{O}_{2}$, $\mathcal{O}_{3}$, and $\mathcal{O}_{4}$, at $\delta=0$ and $\delta=0.05$, respectively. Here, $\mu$ represents the chemical potential. The parameters are $\{t_{0},t_{1},t_{2},t_{3}\}=\{1,0.06,0.2,0.1\}$. For the channel $\mathcal{O}_{i}$, we set $J_{i}=0.1$ and $J_{j\neq i}=0$.}
	\label{fig:OEE}
\end{figure}

{\it \color{blue}Spin-resolved orbital Edelstein effect.---}Spin-orbital magnets carry intertwined spin-orbital textures~\cite{Wang2025soAM}, making the orbital Edelstein effect (OEE) a natural probe of the underlying order even before CPL is applied. In each spin sector, an electric field induces a nonequilibrium orbital polarization, $m_i^s=\alpha_{ij}^sE_j$~\cite{yoda2015current,Yoda2018OEE,niu2019mixed,supplemental}.
The spin-resolved OEE tensor $\alpha_{ij}^{s}$ is therefore directly constrained by the same symmetries that classify the spin-orbital order.

Since orbital polarization is axial whereas the electric field is polar, the linear OEE vanishes in inversion-symmetric systems. Thus, when $[C_{2}\Vert\mathcal{P}]$ relates the two spin sectors, their OEE responses have opposite signs and cancel in the total response, while the spin-resolved OEE (SOEE), $\alpha_{ij}^{S}=\alpha_{ij}^{\uparrow}-\alpha_{ij}^{\downarrow}$, can remain finite. Mirror symmetry further selects the allowed components: a vertical mirror $[E\Vert\mathcal{M}_{j}]$ allows the response to an electric field normal to the mirror plane and forbids that parallel to it. Accordingly, $[C_{2}\Vert\mathcal{M}_{j}]$ makes the two spin sectors display identical OEE for the normal component but opposite OEE for the parallel component. The SOEE in mirror-symmetric spin-orbital magnets is therefore intrinsically anisotropic and directly tied to the underlying symmetry.

These symmetry constraints directly diagnose the pre-driven phases. In {\zb 2D}, only the out-of-plane orbital polarization is well defined, so the relevant tensor elements are $\alpha_{zx}^{s}$ and $\alpha_{zy}^{s}$. For $\delta=0$ [Fig.~\ref{fig:OEE}(b)], the $\mathcal{O}_{4}$ phase retains inversion symmetry and therefore exhibits no OEE, whereas the $\mathcal{O}_{2}$ and $\mathcal{O}_{3}$ phases allow finite SOEE, with $\alpha_{zx}^{S}\neq0$ and $\alpha_{zy}^{S}\neq0$, respectively. Once $\delta\neq0$ is introduced [Fig.~\ref{fig:OEE}(c)], inversion symmetry is broken and finite OEE becomes allowed in the $\mathcal{O}_{4}$ phase as well. The $\mathcal{O}_{3}$ and $\mathcal{O}_{4}$ phases then share the same symmetry and response structure: $[C_{2}\Vert\mathcal{M}_{x}]$ enforces $\alpha_{zx}^{\uparrow}=\alpha_{zx}^{\downarrow}$ while allowing $\alpha_{zy}^{S}\neq0$. In contrast, the $\mathcal{O}_{2}$ phase at finite $\delta$ and the $\mathcal{O}_{5}$ phase are ferrimagnetic, so their spin-resolved OEE responses are no longer symmetry related. The $\mathcal{O}_{1}$ and $\mathcal{O}_{6}$ phases retain $C_{3}$ symmetry and therefore exhibit no OEE.

Under CPL, the real-space mirror symmetries are broken, but their combinations with time reversal can remain preserved. Since time reversal itself does not forbid the OEE, the symmetry-allowed tensor structure of the response remains unchanged by the drive. The SOEE thus provides a complementary probe of the underlying spin-orbital order both before and after driving, and thereby of the parent order responsible for the even-, odd-, and, in particular, mixed-parity altermagnetic states.

{\it\color{blue}Discussions and conclusions.---}In summary, we have shown that mixed-parity altermagnetism constitutes a distinct altermagnetic phase beyond the even- and odd-parity classes. Spin-orbital magnets provides a particularly favorable platform for realizing this phase, because magnetic compensation can occur locally on each site through orbital-selective coupling to opposite spins. We further showed that the characteristic spin-resolved orbital Edelstein effects provide {\zb an} electrical probe for the parent spin-orbital orders of the mixed-parity phase.

From an experimental perspective, layered materials and magnetic heterostructures with inequivalent magnetic layers provide promising platforms. Possible candidates include layered vanadates, Fe-based layered compounds, and double perovskites such as Ba$_{2}$CaOsO$_{6}$~\cite{Rodrigo2025}, Ba$_{2}$YReO$_{6}$, and Sr$_{2}$MgOsO$_{6}$. In these systems, a gate-controlled potential difference can play the role of the staggered potential $\delta$, while CPL provides a tunable knob for driving parity conversion. Since the zero-magnetization condition in spin-orbital magnets does not rely on equivalent sublattices, the material constraints are less restrictive than in altermagnets with conventional Néel order. Our results therefore extend altermagnetism beyond the even- and odd-parity cases to the mixed-parity regime, identifying mixed-parity altermagnetism as an intermediate spin-split phase connecting them.

{\it\color{blue}Acknowledgements.---} Z.-Y. Zhuang would like to thank Xiao-Jiao Wang and Shu-Xuan Wang for fruitful discussions.
Z.-Y.Z. and Z.Y. are supported by
Guangdong Basic and Applied Basic Research Foundation (Grant No. 2023B1515040023), 
and Fundamental and Interdisciplinary Disciplines 
Breakthrough Plan of the Ministry of Education of China (JYB2025XDXM403).
S.B.Z. is supported by the National Natural Science Foundation of China (Grant No.~12488101), the Innovation Program for Quantum Science and Technology (Grant No. 2021ZD0302801).
L.H.H. is supported by National Key R\&D Program of China (Grant No. 2025YFA1411501), the National Natural Science Foundation of China (Grant Nos. 12561160109, 12574148), the Fundamental Research Funds for the Central Universities (Grant No. 226-2024-00068).

\bibliography{dirac.bib}

\end{document}